\documentstyle[psfig,aps]{revtex}

\begin{document}
\draft

\title{Deconvolving the information from an imperfect spherical 
gravitational wave antenna}

\author{Stephen M.  Merkowitz\thanks{Present affiliation INFN-Laboratori 
Nazionali di Frascati, Via Enrico Fermi 40, I-00044 Frascati (Roma), ITALY} 
and Warren W.  Johnson}

\address{Department of Physics and Astronomy, Louisiana State University, 
Baton Rouge, Louisiana 70803}

\date{January 9, 1998}

\maketitle

\begin{abstract}
We have studied the effects of imperfections in spherical gravitational 
wave antenna on our ability to properly interpret the data it will produce.  
The results of a numerical simulation are reported that quantitatively 
describe the systematic errors resulting from imperfections in various 
components of the antenna.  In addition, the results of measurements on a 
room-temperature prototype are presented that verify it is possible to 
accurately deconvolve the data in practice.
\end{abstract}


A spherical gravitational wave antenna was first proposed by Forward in 
1971~\cite{Forward_GRG_1971}.  The large cross section and the ability to 
determine the direction of a gravitational wave have been known for some 
time~\cite{Wagoner_Pavia_1976}, but it was not until recently that its 
potential advantages have been actively 
studied~\cite{Johnson_PRL_1993,Coccia_PRD_1995,Harry_PRD_1996}.  For the 
past few years, several groups have been researching the possibility of 
constructing large scale spherical antennas~\cite{Omni_1997}.

To realize the full potential of these detectors, we must first be able to 
understand their behavior and properly interpret the data they will 
produce.  The technique needed will depend on the particular type and 
arrangement of sensors used to detect the motions of the sphere.  We have 
proposed and investigated the ``truncated icosahedral'' (TI) 
arrangement~\cite{Johnson_PRL_1993} for six identical radial motion 
resonant transducers.  We have shown that this arrangement on a perfect 
sphere, but with equally noisy transducers, has the property that all five 
quadrupole components of the gravitational strain tensor are measured with 
equal sensitivity~\cite{Merkowitz_PRD_1995}.  Other arrangements of 
transducers have been proposed~\cite{Lobo_EPL_1996,Zhou_PRD_1995}, but most 
of them do not share this property~\cite{Stevenson_PRD_1997}.

The TI model, also called a truncated icosahedral gravitational wave 
antenna (TIGA), has been criticized on the basis that normal (small) 
imperfections will break the perfect symmetry, possibly having drastic 
effects on such a degenerate system, thereby making it difficult to extract 
the desired information from the data.  We report here the theoretical and 
experimental results of an investigation into the effects of imperfections 
in a spherical antenna that refutes these criticisms and shows that the 
symmetry breaking can be easily handled.

The normal modes of an uncoupled elastic sphere can be described by the 
spacial eigenfunctions $\vec{\Psi}_{n\ell m}\left({r,\theta,\phi}\right)$ 
(functions of the spherical harmonics) and the purely time dependent mode 
amplitudes $a_{n\ell 
m}(t)$~\cite{Merkowitz_PRD_1995,Ashby_PRD_1975,Lobo_PRD_1995}.  In general 
relativity only the $\ell=2$ quadrupole modes strongly interact with a 
gravitational wave so for the remainder of this discussion we will assume 
$n=1$, $\ell=2$ and drop these subscripts from our notation.  We denote the 
orientation of the 5 quadrupole modes relative to the lab frame by 2 Euler 
angles $\beta_m$ and $\gamma_m$.  If the quadrupole modes are degenerate 
these angles are arbitrary (restricted only by orthogonality of the modes).  
Imperfections of the sphere can lift the degeneracy causing the modes to 
``fix'' themselves in a particular orientation.  This effect has been shown 
to be the main result of symmetry breaking in an {\it uncoupled\/} sphere; 
the ``shape'' of the quadrupole mode remains mostly 
unchanged~\cite{Merkowitz_PRD_1996,Coccia_PLA_1996}.

We consider that a number of small resonant transducers, with index $j = 1 
\ldots J$, are added to the surface of the sphere to mechanically transform 
the small motion of the surface into a large motion of the resonator.  We 
assume the resonators are constructed to obey a one-dimensional harmonic 
oscillator equation.  The particular arrangement of resonators is indicated 
by their locations $\left({\theta_j,\phi_j}\right)$ on the sphere surface 
and their coupling to the surface motion in different directions  
$\vec{\epsilon}_j$.  In the case of ideal radial resonators we would set 
$\vec{\epsilon}_j=\hat{r}$.  The relative amount of surface displacement 
coupled to each resonator for a particular mode is specified by a ``pattern 
matrix'' defined as
\begin{equation}
	B_{mj} \equiv
	\frac{1}{\alpha} \vec{\epsilon}_j \cdot 
	\vec{\Psi}_m\left({R,\theta_j,\phi_j}\right),
\end{equation}
where $\alpha$ depends on the material and the radius $R$ of the sphere 
\cite{Wagoner_Pavia_1976,Merkowitz_PRD_1995}.  Again, we assume that the 
transducers strongly interact only with the quadrupole modes of the sphere, 
which is appropriate if the mass of a resonator is small compared to the 
mass of the sphere~\cite{Lobo_EPL_1996,Lobo_PRD_1998}.

The complete set of coupled equations of motion for the sphere and 
resonators can now be written~\cite{Merkowitz_PRD_1997}
\begin{equation}
	\left[{ \begin{array}{cc}
		\bbox{M}^s & 
		\bbox{0} \\
		\alpha 
		\bbox{M}^r 
		\bbox{B}^T & 
		\bbox{M}^r
	\end{array} }\right]
	\left[ \begin{array}{c}
		\bbox{\ddot a}(t) \\
		\bbox{\ddot q}(t)
	\end{array} \right] +
	\left[ \begin{array}{cc}
		\bbox{K}^s & 
		-\alpha \bbox{B} \, 
		\bbox{K}^r \\
		\bbox{0} & 
		\bbox{K}^r
	\end{array} \right]
	\left[ \begin{array}{c}
		\bbox{a}(t) \\
		\bbox{q}(t)
	\end{array} \right]
	= 
	\left[ \begin{array}{cc}
		\bbox{I} & 
		-\alpha 
		\bbox{B} \\
		\bbox{0} & 
		\bbox{I}
	\end{array} \right]
	\bbox{f}(t),
\label{eqn:eom_matrix}
\end{equation}
where we have defined 4 diagonal matrices whose elements are the mass $m$ 
and spring constants $k$ of both the sphere modes (superscript $s$) and the 
resonators (superscript $r$),
\begin{equation}
	M^s_{jm} \equiv \delta_{jm} m^s_m, \;\;
	M^r_{jm} \equiv \delta_{jm} m^r_j, \;\;
	K^s_{jm} \equiv \delta_{jm} k^s_m, \;\;
	K^r_{jm} \equiv \delta_{jm} k^r_j.
\end{equation}
The column vector $\bbox{a}$ has five components, one for each sphere mode 
amplitude, and has a one-to-one correspondence with the quadrupole 
components of the gravitational wave strain 
tensor~\cite{Merkowitz_PRD_1995,Lobo_PRD_1995}.  The column vector 
$\bbox{q}$ has $J$ components, one for each resonator displacement, and are 
the observable quantity.  The column vector $\bbox{f}$ contains any forces 
(such as a gravitational wave) acting on the sphere and resonators.

A normal mode solution of these coupled equations is possible using 
standard techniques~\cite{Merkowitz_PRD_1997}.  We find the response of the 
normal modes $\bbox{\eta}(\omega)$ to any external forces, including 
gravitational waves, is given by
\begin{equation}
	\bbox{\eta}(\omega) 
	= 
	\bbox{G}(\omega) 
	\bbox{V}^{-1} 
		\left[{ \begin{array}{cc}
		\bbox{M}^s & 
		\bbox{0} \\
		\alpha 
		\bbox{M}^r 
		\bbox{B}^T & 
		\bbox{M}^r
	\end{array} }\right]^{-1} 
	\left[ \begin{array}{cc}
		\bbox{I} & 
		-\alpha \bbox{B} \\
		\bbox{0} & 
		\bbox{I}
	\end{array} \right] \, 
	\bbox{f}(\omega).
	\label{eqn:nc_solution}
\end{equation}
The matrix $\bbox{V}$ is constructed from the eigenvectors of the normal 
modes, and the matrix $\bbox{G}(\omega)$ is constructed from their 
eigenvalues.  The matrix $\bbox{V}$ can also be used to return to the 
original sphere and resonator coordinates
\begin{equation}
   \left[\begin{array}{c}
      \bbox{a}(\omega)  \\
      \bbox{q}(\omega)
   \end{array} \right]
	= 
   \bbox{V} \, \bbox{\eta}(\omega).
   \label{eqn:V_eta}
\end{equation}
$\bbox{V}$ is always invertible thus it is possible to use the inverse of 
Eq.~(\ref{eqn:V_eta}) to transform the data to normal coordinates where the 
frequency response is simple.

Examining Eqs.~(\ref{eqn:nc_solution}) and~(\ref{eqn:V_eta}), we see that 
in order to obtain all the information about an excitation we must have 
knowledge of both the sphere $\bbox{a}$ and resonator $\bbox{q}$ modes even 
though only the resonator motion is observable.  A simple solution emerges 
if we limit ourselves to the case of ``close'' to the TI arrangement (six 
radial resonators placed at the center of six non-antipodal pentagon faces 
of an imaginary TI concentric to the sphere).  The eigenfunctions of an 
ideal TIGA are such that the motion of the resonators mimic the ellipsoidal 
(quadrupole) deformation of the sphere's surface either in phase or 
anti-phase~\cite{Merkowitz_PRD_1995}.  We can picture the collective motion 
of the six resonators to describe six ``resonator ellipsoids'', five of 
which are mimicking the ``quadrupole ellipsoids'' of the sphere (the sixth 
resonator ellipsoid is just a ``breathing'' sphere: the six resonators are 
moving in unison with equal amplitude and the sphere surface does not move 
at all).  Each individual resonator displacement $q_j$ represents a 
superposition of the point radial deformation of the six resonator 
ellipsoids at a particular location.  The transformation between point 
radial deformations $\bbox{q}$ and ellipsoidal amplitudes $\bbox{g}$ is 
given by the pattern matrix $\bbox{B}$:
\begin{equation}
	\bbox{g} = \bbox{B} \; \bbox{q}.
	\label{eqn:pattern_matrix}
\end{equation}
The resonator ellipsoids $\bbox{g}$ are proportional to the sphere modes 
$\bbox{a}$ so we sometimes refer to them as ``mode channels.'' We also 
point out that Eq.~(\ref{eqn:pattern_matrix}) can be seen as a linear 
coordinate transformation between the measured point radial displacement 
and a 5-dimensional abstract vector space based on the spherical harmonics 
that has a one-to-one correspondence with the five quadrupole components of 
the gravitational strain tensor~\cite{Merkowitz_PRD_1995,Lobo_PRD_1998}.  
While not unique to TIGA, such a simple transformation is not possible for 
most other arrangements of transducers.

We are now prepared to discuss how imperfections effect our ability to 
deconvolve the signal from the transducers.  One source of error is the 
resonator ellipsoids may be deformed so that they no longer mimic the 
sphere surface.  This could lead to errors in calculating both the matrix 
$\bbox{V}$ and the mode channels $\bbox{g}$.  Another source of error is 
the uncertainty of the values of all the system parameters.  Most of the 
parameters can be measured, either directly or indirectly, but we will be 
limited by the accuracy of that measurement.

The transformation matrix $\bbox{V}$ can be measured by applying a 
continuous sinusoidal force anywhere on the sphere's surface at the 
frequency of one of the normal modes.  (Note that this technique requires 
the normal modes to be {\it non-degenerate\/}, thus it is actually 
preferable to have a small amount of symmetry breaking.) The frequency 
response of the resonators will be simple because they are being driven at 
a single frequency.  From Eq.~(\ref{eqn:V_eta}) we see that the amplitude 
and phase of their response make up a single column of $\bbox{V}$.  By 
exciting each normal mode in turn, the complete $\bbox{V}$ matrix can be 
measured.  The only assumption made in calculating $\bbox{V}$ is that the 
resonators are ``close'' to the TI arrangement so that an ideal pattern 
matrix $\bbox{B}$ can be used and the sphere modes $\bbox{a}$ can be 
replaced by the mode channels $\bbox{g}$.

The quadrupole mode orientation angles $\beta_m$ and $\gamma_m$ can be 
measured directly only before the resonant transducers are 
attached~\cite{Merkowitz_PRD_1996}.  This is also true for the the mass of 
the resonators $m^r_j$ and the sphere modes $m^s_m$, and their respective 
spring constants $k^r_j$ and $k^s_m$.   Once the system is coupled 
together these parameters cannot be measured independently.  The positions 
of the resonators $\left({\theta_j,\phi_j}\right)$ can be easily measured, 
but precise values of their couplings $\vec{\epsilon}_j$ are much harder to 
determine.

We developed a numerical simulation of a TIGA that calculates the error 
associated with a measurement due to imperfections and asymmetries as well 
as our uncertainty of the precise values of the above parameters.  We 
simulated a gravitational wave burst excitation of an imperfect TIGA and 
used the above analysis techniques to estimate the direction of the wave 
from the simulated output of the transducers.  The analysis assumed 
parameter values of an ideal TIGA and a measured $\bbox{V}$ matrix.  The 
``real'' (perturbed) parameters of the imperfect TIGA were calculated by 
adding uniformly distributed random numbers to the ideal values.  The range 
of values of the random number was a fraction (tolerance) of the ideal 
value ($2\pi$ for the angles).

\begin{figure}
\centerline{\psfig{figure=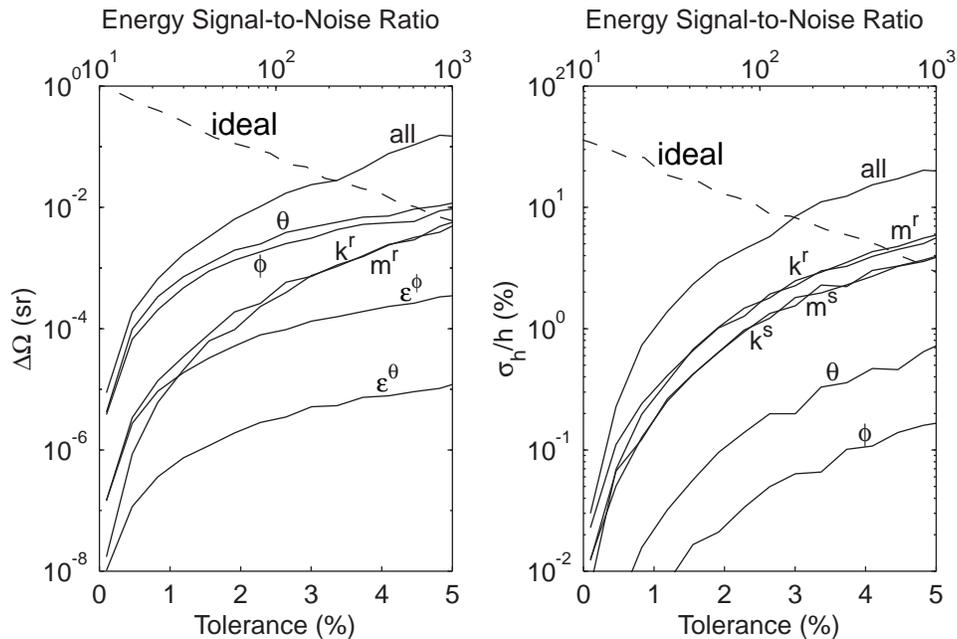,width=5.0in}}

\caption{The solid angle direction estimation error and the percentage 
error in estimating the strain magnitude as functions of the tolerance on 
the system parameters of a TIGA and as functions of the signal-to-noise 
ratio for an ideal TIGA.  Each solid line corresponds to the bottom axis 
and represents the results of a 200 trial simulation with the indicated 
parameter varied within the corresponding tolerances.  The dashed lines 
corresponds to the top axis and represents the results of a simulation of 
an ideal TIGA for a range of signal-to-noise ratios.}

\label{fig:direction_error}
\end{figure}

The results of the simulation are shown in Fig.~\ref{fig:direction_error}.  
Plotted is the solid angle estimation error $\Delta\Omega$ and the 
percentage error $\sigma_h/h$ for estimating the strain amplitude for a 
range of tolerances.  Parameters not shown had errors close or equal to 
zero.  Because the tolerance is expressed as a percent of the ideal 
parameter values, varying the magnitude of the ideal parameters (for 
example doubling the mass of the sphere) produced identical results.  Also 
varying the direction and polarization amplitudes of the gravitational wave 
produced similar results.  From these results we conclude that the analysis 
becomes unreliable only after the tolerance of all the parameters exceeds 
about 3\%.  This is certainly an obtainable level of precision; one expects 
to set the tolerances on a real antenna to better than 1\%.

Also shown in Fig.~\ref{fig:direction_error} are the errors due to a finite 
signal-to-noise ratio, calculated in a similar fashion to the analysis of 
Zhou and Michelson~\cite{Zhou_PRD_1995}, but using the direction finding 
algorithm used in our measurements~\cite{Merkowitz_PRD_1996}.  Note that 
this is a random error whereas the tolerance errors are systematic.  
Comparing the two sources of error, we find that one would need a 
signal-to-noise ratio of about 1000 in energy before the tolerance errors 
become significant.  While one might hope to observe sources at this level, 
the most optimistic predictions lead to considerably smaller 
values~\cite{Finn_OMNI_1997}.

To further evaluate the feasibility of a spherical antenna, we constructed 
a 0.8~m diameter, Aluminum alloy, room-temperature prototype 
TIGA~\cite{Merkowitz_PRD_1996,Merkowitz_PRD_1997}.  Six radial resonant 
transducers were attached to the surface in the TI arrangement: placed at 
the center of each of the six lower pentagon faces.  The response of each 
resonator was converted to an electrical signal by a piezoelectric strain 
gauge, whose output was recorded on a high speed data acquisition system.  
The apparatus is described in detail in Ref.~\cite{Merkowitz_Thesis}.

Since no laboratory source of gravitational waves exists, we chose to apply 
an impulsive force to the surface of the prototype to test the developed 
techniques.  To quantify the test, we checked if the location of the 
impulse could be measured from the response of the resonators.

\begin{figure}
\centerline{\psfig{figure=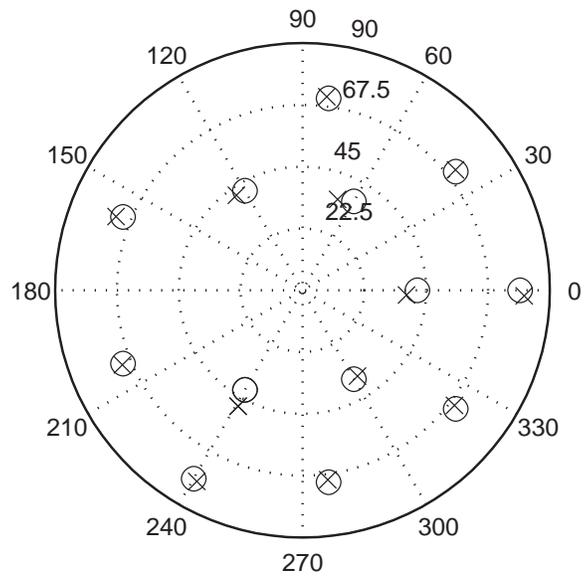,width=3.0in}}

\caption{Location of several impulses applied to the prototype TIGA.  Each 
x marks the location calculated from the motion sensor data and the nearby 
o marks the location of the center of the shaker measured geometrically.}

\label{fig:burst_location}
\end{figure}

We began by measuring the transformation matrix $\bbox{V}$ as described 
above.  Impulsive forces were then applied by sending a short electrical 
pulse to a piezoelectric shaker attached to the surface of the TI.  The 
data from the transducers were recorded, transformed to normal coordinates, 
fit for phase and amplitude, and finally transformed to mode channels.  
From the mode channels the location of the impulse was calculated.  The 
algorithm used to determine the location of the impulse can also be used to 
calculate the direction of a gravitational 
wave~\cite{Merkowitz_PRD_1996,Merkowitz_PRD_1997}.  The results of this 
analysis for the various impulse locations is shown in 
Fig.~\ref{fig:burst_location}.  From these results, we conclude that it was 
possible to reconstruct the location of the impulses, within the accuracy 
of the experiment, without major difficulties.

In summary, from the numerical and experimental results we conclude that 
imperfections in a spherical antenna can be easily handled and that the 
error associated with the asymmetries are small.  We find that the TI 
arrangement is fairly robust to these normal imperfections.  The techniques 
for deconvolving the data are mostly linear algebra, making their 
implementation simple in an automated data analysis system.  The 
measurement of the transformation matrix $\bbox{V}$ takes into account most 
deviations from perfect symmetry and enables the data to be transformed to 
a space where the frequency complications can be easily handled.  The 
algorithms were tested on the prototype TIGA and found to be consistent 
with the measured results.  Finally, since all the techniques described and 
tested can be applied {\it in situ\/}, they are directly applicable for use 
on a real spherical antenna searching for gravitational waves.

We thank W.~O.~Hamilton for essential advice and support.  This research 
was supported by the National Science Foundation under Grant No.  
\mbox{PHY-9311731}.


\end{document}